# Intervenability as a Design Requirement for Autonomy and Oversight within Human-Centered AI

Thomas Herrmann [0000-0002-9270-4501]

Information and Technology Management, IAW, Ruhr-Universität Bochum, Bochum, Germany
thomas.herrmann@rub.de

**Abstract.** Based on the literature and several practical examples of possible AI applications, we outline the concept of intervenability. This new phenomenon is not covered by emergency shutdowns, workarounds, or the reconfiguration of automated systems. Intervenability instantiates the principles of controllability, autonomy, oversight, and keeping humans in the loop in the context of AI. We provide a taxonomy that encompasses a range of possibilities for intervening activities and differentiates them regarding the mental effort of the users. This taxonomy extends the scope of interventions from real-time control of automated processes to AI-based discrete case-related decision-making. This is in accordance with human-centered AI, which seeks to combine human strengths with the usage of AI. We demonstrate how intervenability can potentially contribute to the ongoing development of human capabilities on the one hand and to further technical improvement by reconfiguration of AI on the other. Exploring and collaboratively reflecting on the effects of interventions as an integral part of organizational practices is key to enabling this continuous improvement on both sides. Intervenability also provides further momentum for the design of an AI that can help realize interventions on its own and advance a smooth transition from intervention to reconfiguration of the AI.

## 1 Introduction

A wide range of risks have been associated with the use of artificial intelligence. The superannuation of jobs, the impairment of human autonomy and freedom of choice, unintended damage from bad-quality AI, and bias in AI decisions that disadvantage certain social groups are prominent examples. The risks are particularly high when AI is directly coupled with the physical environment, such as in mobility, manufacturing, or weapons applications, or whenever decisions made by AI are implemented automatically without checks. One contribution to mitigating these risks is to provide opportunities for people to intervene when dealing with AI. DFKI[1] expert Paul Lukowicz said in an interview on the German newscast *Tagesschau* on June 14, 2023: 'It is always problematic when AI does things not subject to human intervention or review.'

In our opinion, the powers of review and intervention are inseparable, and reviewing is a precondition for continuous human control. One has to be able to react if one sees that something is going wrong. However, designing these possibilities for intervening is not trivial. Schmidt and Herrmann [1] propose intervention interfaces between humans on the one side and automated processes or automated decision-making on the other. Implementing intervention possibilities is quite complex, as we will explain with respect to intervention interfaces. It requires measures at the level of the AI itself and the technical infrastructure surrounding it, and at the level of organizational practice.

The implementation of intervenability is thus a socio-technical challenge, and we will answer the questions of what kind of possible activities should be covered by intervenability and how intervention can be realized to be successful. Intervenability can be considered as a variant of controllability [2] and corresponds with approaches that argue for keeping humans in the loop when running automated systems or decision-making [3], [4]. Within the HCAI discourse, controllability is mostly related to human agency and oversight [5] and includes the human-in-the-loop approach that maintains humans' capability to intervene in every decision cycle of the system. Georgieva et al. [6, p. 706] provide an overview of 13 categories of oversight; it includes, for example, a stop button that can be used to cancel a process. However, the available descriptions of controllability do not sufficiently guide the design of human-AI interaction. The relevance of such an additional variant of controllability is related to typical characteristics of AI: it is designed with the goal of minimizing the interactive contributions of those who benefit from employing AI for certain tasks. The final consequence is – at least in certain phases – interaction-free usage of autonomous systems or automated processes. Thus, the EU guidelines [5] suggest that interventions in every decision cycle may be neither possible nor desirable. If controllability is meant as the ability 'to maintain direction over the whole course of the interaction' [2, p. 256], we need a concept that accounts for the diminishing of the need for interaction-driven usage of systems. Furthermore, the goal must be that users do not have to monitor AI in detail and do not have to routinely intervene over and over again. If that were necessary, it would be more appropriate to replace or reconfigure the AI

[1] DFKI: Deutsches Forschungszentrum für Künstliche Intelligenz, German Research Center for Artificial Intelligence

solution. Rather, monitoring should only be required on specific occasions or at specific times, and intervention should be an exception that does not completely terminate a process – like an emergency stop button – but preserves the possibility of AI resuming its activity.

This means that it is not desirable for users to be made permanently suspicious of AI. This kind of trust deficit ('under-trust') would result in an unnecessary control effort and extra human workload – and also a deterioration of the performance of the overall socio-technical system [7], [8]. At the same time, humans must not place too much trust in AI because such 'over-trust', when misplaced, entails serious and costly risks to efficiency and safety, endangering both equipment and lives [7]. The problems of under-trust and over-trust are extensively analyzed by Okamura and Yamada [8], who focus on trust calibration as a process where users appropriately adjust their level of trust to the actual reliability of an AI system. 'Over-trust is poorly calibrated trust that exceeds the reliability of an AI system … Under-trust is poorly calibrated trust that falls short of the AI's reliability …" [8, p. 220335]. Calibration can also be supported by intervention, which not only corrects AI results or processes but also allows AI to be explored and experimented with, triggering critical reflection [9].

In what follows, we describe the need for and the description of intervenability based on the literature and several examples of AI usage. This description will result in a taxonomy of types of intervention. Based on this taxonomy, we can outline how intervention differs from other concepts (Section 3). Based on the elaborated understanding of the concept of intervention, we can describe its interplay with the continuous improvement of AI and the ongoing development of human competencies. The implementation of intervenability has to fulfill certain requirements for interaction design (Section 5) and its embedding into organizational practices (Section 6).

## 2 Background

This section provides an overview of how an understanding of intervention in the context of using AI has developed. By contrast to this context, intervention originally refers to measures that influence human behavior, not that of artifacts: "To intervene is to enter into an ongoing system of relationship, to come between or among persons, groups, or objects for the purpose of helping them. [10, p. 15]" In the context of using technical systems, intervention is more about altering the plan of how usage runs [11]. This plan might be projected on the processes of use by the designer, who has certain models of usage in mind. These models can be implicit or explicit, such as GOMS[2]-models [12]. When applying AI, use models are focused mostly on reducing the need to interact with it.

[2] GOMS: Goals, Objects, Methods, Selection Rules. https://en.wikipedia.org/wiki/GOMS (Retrieved February 20, 2024)

As we will show, intervention is not intended to create sustaining effects on the characteristics of the AI system itself but is focused on the properties of a situation in which task handling with AI takes place, for example, when a car is steered or when a job applicant is selected. In the following three subsections, we prepare a taxonomy of interventions summarized in Figure 1 (Section 2.4). The examples given below refer to this taxonomy in parentheses.

## 2.1 Intervention and intervention interfaces

Schmidt and Herrmann [1] define intervening interaction as an activity of users that temporarily changes the behavior of an automated process. To illustrate, Schmidt and Herrmann draw on the process of parking a car. In a conventional car, users park the car by themselves and exert continuous and fine-grained control. Every 100-500 milliseconds, one evaluates the situation, controls or corrects the steering, and accelerates or brakes. In a car with automatic parking, one controls the process only to the extent that one selects the parking space and starts the parking process. The fine-grained control of steering and moving back and forth is automated – one does not have to be sitting in the car to do it. In an autonomous, potentially driverless car, one moves away from the car, which then independently finds a parking space and parks.

Intervention in a car with automatic parking assistance may mean, for example, that the driver sees a group of preschoolers approaching and stops the automatic parking for a while until the group has passed (Figure 1, A, level 1). That would be the simplest kind of intervention – stop and resume. It is different if a driver realizes, for example, that the car will roll onto the unprotected irrigation area of a tree as it moves forward, and she wants to avoid that. Then, she can take control completely and park by herself, so she takes fine-grained control for a while during this intervention (Figure 1, A, Level 2). However, she could also intervene by only influencing the sensory system by bravely positioning herself as an obstacle between the tree and the car so that changing the direction would be initiated sooner (it would be advantageous if the automatic system recognizes this as human intervention and remembers the obstacle so that one does not have to stay in front of the car all the time). Alternatively, she only influences the decision-making of the automatic system by specifying for driving forward for this particular situation that the change of direction is already made at a greater distance from the obstacle – in this case, the tree (Figure 1, A, Level 3). Or, she only takes over the action of changing direction by pressing a button each time, for example.

Generally speaking, many AI-supported processes usually run in a mode of implicit interaction [13] without an active human contribution. The process takes place according to a plan and depends on the events registered in the AI system's environment. Intervening interaction is an ad hoc change instantiated by extraordinary, unplanned human control, which is effective only within a certain time slot or for a limited area of automated processes (such as the interpretation of sensor data). A systematic intervention user interface comprises several aspects:

- It *enables* a selective perception of automated processes,
- it clearly presents or even suggests possible steps for an intervention to the user,
- and it offers control options with which an intervention can be easily, quickly, and robustly started, completed, terminated, and occasionally revised.

The example of parking already shows that there are various possibilities for implementing intervenability. We define intervention interfaces as interfaces that systematically support all kinds of reasonable intervention activities in an automated system, including preparing and revising undesired effects as well as terminating the intervention. We can already find single possibilities for intervening by actions such as activating the party mode of a heating system or the ad hoc adjustment of water or coffee quantities in a coffee machine for a single cup of coffee. By contrast to these single possibilities, intervention interfaces allow for the use of all types of possible interventions, their integration, and consideration of reasonable points of physical access from where the intervention can be started. If more than one automated process or AI-based decision-making is repeatedly used in the same context, the intervention interface helps to quickly address the process that should be influenced by an intervention.

Intervention can take place with varying degrees of mental effort, from simply switching off the system during a specified timeslot (Figure 1, A, Level 1) to taking over control in fine-grained steps within a specified timeslot (Figure 1, A, Level 2), to temporarily changing data or thresholds (Figure 1, A, Level 3). The invariant characteristic of all types of interventions that we describe here is their *temporal* instantiation – after a certain time span or certain events, the mode that is initiated by the intervention will be terminated, and the automated system will take over the process again. The users will clearly realize whether this resumption has taken place, and they cannot forget to switch back to the regular state or erroneously assume that they did so. This emphasis on temporal change of the system's mode of usage contrasts with the long-lasting effect that might be achieved for the outcome of AI usage in a certain case. What is adapted with a longer-lasting effect is not the AI system itself but the effect of the system on its usage environment.

We speak of *intervenability* if systems have an intervention interface that integrates the variety of possibilities of acting and observing that are required for a successful intervention.

### 2.2 Human-centered AI and advanced intervention

Intervenability into AI is a design option that belongs to a broader research area, namely Human-Centered Artificial Intelligence, or HCAI for short [14], [15]. Various approaches can be assigned to human-centered AI, such as responsible AI [16], keeping the human in the loop [3], collaborative AI [17], [18], hybrid intelligence systems [19] and explainable AI (XAI) [20], [21]. We suggest that HCAI is most clearly characterized by the underlying assumption that hybrid intelligence has "… the ability to accomplish complex goals by combining human and artificial intelligence to collectively achieve superior results than each of them could have done in separation and continuously improve by learning from each other." [19, p. 275]. This definition of HCAI accepts that both sides can continuously evolve by learning, and thus, it differs from a more general notion of human-centered technology, such as that preferred by Shneiderman [14], which does not emphasize the potential autonomy of AI as an agent that can evolve on its own. It is important to consider this autonomy since future users will have to deal with not only one but also manifold AI-driven autonomous processes, as in the smart home [22]. In such situations, users will neither have the capacity nor time to exercise continuous and detailed control and oversight of every process. This is one reason why focusing on occasional intervention is important.

A further aspect of HCAI is the explainability of AI (XAI), which is usually provided by AI itself. It is also indispensable because human intervention and control are only possible if one can identify the extent to which an intervention is possible and useful, where it can be applied, and what effect it will have. One could argue that it is unnecessary to know the algorithm of AI-based parking when one can simply take control so that the car does not drive too close to a tree. On the contrary, we suggest that the driver should know whether the AI cannot perform a certain action or decision or whether something on the car is not working as it should, e.g., if a camera on the vehicle's underside is dirty. If the AI does not account for whether the car is driving into the irrigation area of a tree, then it is important for human intervention to know whether that is because the sensor technology is not in for that situation or whether the system is not interpreting the data correctly.

The original definition of intervention [1] addresses AI-driven real-time processes. Considering the broader scope of AI applications that are covered by HCAI, we suggest that intervenability refers not only to automated physical processes, such as autonomous vehicles but also to workflows of AI-based decision-making for certain types of specific cases. There are AI-assisted processes, such as identifying social fraud [23], that automatically result in consequences for those affected without human decision-makers being able to intervene (yet). In contrast, humans need to be able to intervene in those processes where an AI component specifies a decision that is then executed unquestionably by human actors or by traditional software, and that may result in people with certain characteristics – e.g., dual citizenship – being negatively affected. Humans must be able to veto AI-based decisions (Rakova et al., 2021). HCAI assumes that humans are the better decision-makers as soon as it is no longer just a matter of mastering complex contexts and huge amounts of data but of making decisions when there is ambiguity, as in the context of negotiations, or when only insufficient information is available [24].

Another real-world example of applications that do not resemble the processes involved in automatic vehicle control are enterprise resource planning systems or warehouse systems [25] that include AI components. Such a component can decide, for example, that a certain good is to be reordered. Following this, another AI system can then select the supplier and schedule the transport, or the order can be executed by a conventional IT infrastructure or by human clerks. No matter which of the three options is realized, before proceeding, a human decision-maker must be able to review the AI-based order and to intervene if necessary and adjust the order timing or quantity of goods for the specific situation (Figure 1, B, Level 2). A similar situation applies – as another example application area – to industrial plants where AI automatically provides maintenance instructions [26], [27]. Based on sensors, it could be determined, for example, that a pump needs to be replaced due to pressure fluctuations. Before the usual process is started with a repair order, human decision-makers must check whether the pump really might fail and whether the costs associated with the replacement are justified. Especially when such predictive maintenance systems are introduced, there are often false alarms in the beginning that would lead to unnecessary repairs if they were all taken seriously. Intervention must therefore be enabled for holistic, socio-technical processes in which AI, conventional IT, and human actors interact [28], [29].

### 2.3 Intervention as exploration

Intervention is also an opportunity to run experiments or to find answers to what-if questions that both help to explore the AI system and its application domain (Figure 1, Level 4). This can be explained by the example of AI-based predictive maintenance alerts for industrial equipment. Suppose an alert is issued that a pump needs to be replaced due to excessive pressure fluctuations. The plant operator may intentionally ignore this warning because she wants to see if the pump will work for a while longer. She may already have a spare pump ready for quick replacement, but she wants to wait and see if the prediction is valid so she can learn something about the reliability of the AI system. This ignoring of the warning by the plant operator means that she rejects the AI outcome. This rejection is presented as the first level within the taxonomy of intervention possibilities (Figure 1, B, Level 1). The operator might decide to jump to a higher level of intervention by increasing the threshold value up to which, in her opinion, pressure fluctuations are still considered tolerable and not indicative of a defect (Figure 1, B, Level 3). The plant operator could even go further and inspect old cases where correct warnings of pump failures were given. She could then check whether these correct warnings would still have been given under the condition of the increased threshold value. This strategy means that intervention is combined with an exploration of the system (Figure 1, Level 4).

Additionally, or alternatively, the AI system itself could react to a rejection by prompting the user to propose changes in data or rules that lead to more acceptable AI outcomes. In this way, an intervention interface could be extended to include functionality that encourages experimentation and exploration on the part of the users, thus supporting their learning processes. Such a temporary increase of a threshold would also be an intervention that helps to learn something about maintenance requirements in general.

### 2.4 A taxonomy of intervention

| | | | | | | |
|---|---|---|---|---|---|---|
| *A) AI-based continuous real-time control* | **interruption** until a time slot has expired or while user **waits** for a situational change | **interruption** while user **takes over** detailed, fine-grained interactive **control** | user **modifies** data, thresholds or rules to **temporarily influence** the control of a process | user **temporarily modifies** parameters or rules **to explore** the effects of the modification on a process | user **modifies** parameters or rules to **influence** the control of a process for **a time slot in the future** | user **influences one or several** processes to achieve a temporal effect and are **aware of side effects** on other processes |
| | *Increasing mental effort for conducting an intervention* → | | | | | |
| *B) AI-based discrete case related decision making* | **rejection** and **resumption** of decision making when new information is available or after a deadline | **rejection** and/ or **modification** of a decision made for the current case | user **modifies** data, thresholds or rules **to influence** decision making **for the current case** | user **modifies** parameters or rules **to explore** the effects of the modification on **the current case** | user **modifies** parameters or rules to **influence** decision making **for certain types of future case** | user **influences** decision making for **several interdependent cases** and are aware of side effects on decisions |
| | *Level 1* | *Level 2* | *Level 3* | *Level 4* | *Level 5* | *Level 6* |

**Fig. 1.** A taxonomy of intervention activities

We summarize the background section by the taxonomy presented in Figure 1. Two types of AI applications are presented. The upper part of the diagram presents interventions into AI-based continuous real-time control such as using autonomous cars, control of airplanes, agricultural machinery, and smart homes; the lower part of the diagram presents interventions into AI-based decision making such as fraud detection, handling applications for reimbursement in insurance companies and ordering of goods by warehouse systems. AI-based predictive maintenance is an example that demonstrates that there are applications between these two types since the continuous monitoring of machinery is a real-time process, while every warning can be considered as a case to be handled by the plant operator. The possibilities that represent interventions are ordered from the left to the right side with respect to the level of mental effort required. Table 1 presents examples of the diverse levels for both types of AI applications. The mental effort here refers to the activities to be carried out, the observation of effects that are caused by intervention, or the anticipation of the effects of an intervention. We consider the degree of planning required to be decisive for the level of mental effort caused by an intervention.

This assumption is based on action regulation theory [30]. It distinguishes between the development or selection of goals, orientation, planning, monitoring of action and outcome, and feedback processing. Action regulation theory differs from other theories that analyze concepts of task-related complexity and mental effort but do not regard the relevance of planning [31]. We suggest that goal setting and orientation occur before the intervention. They lead to the decision to instantiate an intervention. The diverse levels of mental effort in our taxonomy refer to the extent of planning and feedback-based adjustments to plans that are required. Level 1 (Figure 1) mainly requires monitoring and a single planning decision about when to resume AI-based task handling, representing the lowest mental effort.

**Table 1.** Examples of diverse levels of mental effort for conducting interventions

| Level | **AI-based continuous real-time control or surveillance** (examples for autonomous driving and smart homes) | **AI-based discrete, case-related decision-making** (examples for warehouse systems and related processes of purchasing and delivery) |
|---|---|---|
| 1 | Parking is interrupted until a group of children has passed; the brake assistant is interrupted. The readiness of an alarm system is interrupted for a brief time. | A single order of goods is delayed in an AI-based warehouse system until a storm has passed. |
| 2 | The driver takes over control until a construction site has been passed. | The user enters data into the warehouse system that reflects specific requirements caused by an emergency. |
| 3 | Speed is reduced while driving through a scenic landscape. | The user enters rules according to which the delivery of certain goods can no longer be classified as a matter of priority handling. |
| 4 | A threshold is raised to see whether a predictive alarm is appropriately avoided, although the water level of the heating system fluctuates. | A route planner avoids a specific route to see whether this allows for faster transport during rush hours. |
| 5 | Reducing the watering system for a period of days. | Delaying the handling of certain types of orders until a supplier has updated some relevant warranties. |
| 6 | Delaying several energy-consuming processes in a smart home until the sun is back | Putting a certain order on hold until a court precedent is decided, although these decisions are needed to proceed with other delivery processes. |

Detailed, interactive control during an intervention or modification of AI outcome may rely on existing planning patterns and include unconscious routines, thus requiring slightly increased mental effort (Level 2). Just prompting the system with short commands – such as 'slow down' or 'consider the availability of staff' – are examples for level 2. Altering data, rules, or thresholds that influence the AI system requires some anticipation of the modifications' effects and thus represents the next higher degree of effort (Level 3). As soon as users cannot be sure whether the applied plan-ning of Level 2 or 3 will be successful, the consideration of feedback and adaptation of plans becomes relevant (Level 4). Level 4 represents a situation where users' interventions are similar to trials or experiments, thus representing a mode of exploration. Rejecting an AI outcome is a simple act but is more challenging if the effects of this rejection need to be observed and are a matter for further adaptation of the system. The more the effects of interventions become apparent in the future only, i.e., the less direct the feedback will be, the more planning and anticipation are required to achieve desired results (Level 5). If a person employs and monitors not only one but several AI-based processes, side effects between the processes may have to be considered, as occurs in the smart home or when the same customer is affected by several decision-making cases. In these interdependent processes or cases, exploring or future assessment of intervention effects becomes even more mentally challenging (Level 6).

We have derived the distinct levels of the taxonomy in Figure 1 by analytical consideration in relation to action regulation theory but not by empirical investigations of our own.

## 3 Intervention – an alternative among other options

### 3.1 Intervention versus emergency stop

Introducing a new principle or design criterion, such as intervenability, raises the question of whether such an additional option is necessary alongside existing criteria or options for user interface design. For example, the conventional emergency stop button could be suggested already as a metaphor for how systemic intervenability could also be realized (see Figure 2).

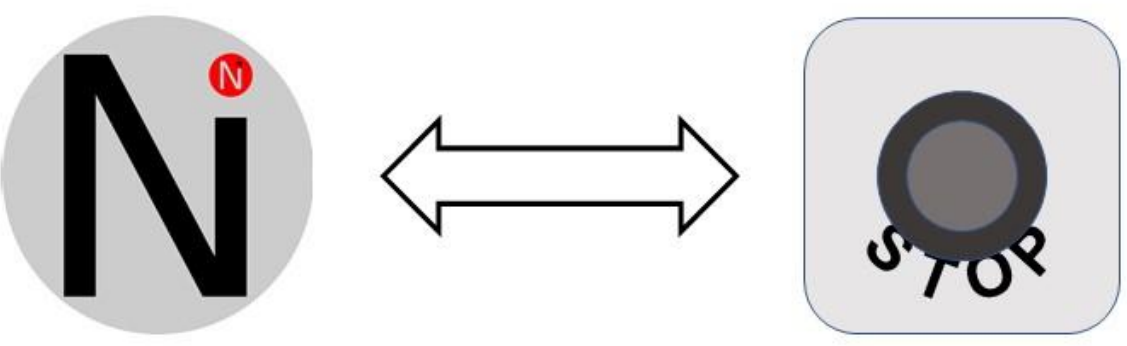

**Fig. 2.** Intervention versus emergency stop

However, it is precisely by distinguishing it from the emergency stop button that one can explain the special nature of the intervention. Take, for example, the brake assistant in trucks, which is supposed to prevent serious rear-end collisions. Cars might get too close in front of a truck coming via highway access, and the assistant triggers an unwanted braking process [32]. Some drivers, therefore, switch off the assistant in such situations. However, they occasionally fail to switch it on again. This again increases the risk of accidents. An intervention button (see Figure 2), on the other hand, would only trigger a temporary interruption of the automatic system, with the assistant being switched on again after a preconfigured time and signaling this accordingly [33]. Interventions are, therefore, always limited and do not permanently interrupt the regular process. While the resumption after using an emergency button requires the typical – possibly awkward – restart procedure, an intervention interface allows for triggering an immediate resumption either by the human agent or the system itself. Thus, in contrast with the emergency button, the system is never shut down completely but is still active in the background, ready to take over the task handling again.

Since using the emergency stop button might cause a costly interruption and a time-consuming system restart, people tend to avoid using it and prefer to let a current automated process run as it is. This might cause so-called workarounds [34]. Workarounds are specific ways of intentionally handling a specific task, although these ways are not planned, expected, or allowed; in some cases, they may even be illegal. Usually, workarounds do not aim to modify or explore the technical system but alter organizational procedures that surround technical artifacts. An intervention button is part of an intervention interface, and with the same button or controls combined, the resumption can also be started. Consequently, intervention is a third option between complete termination on the one side and letting the system run as it is, including possible workarounds, on the other side (see Fig 3).

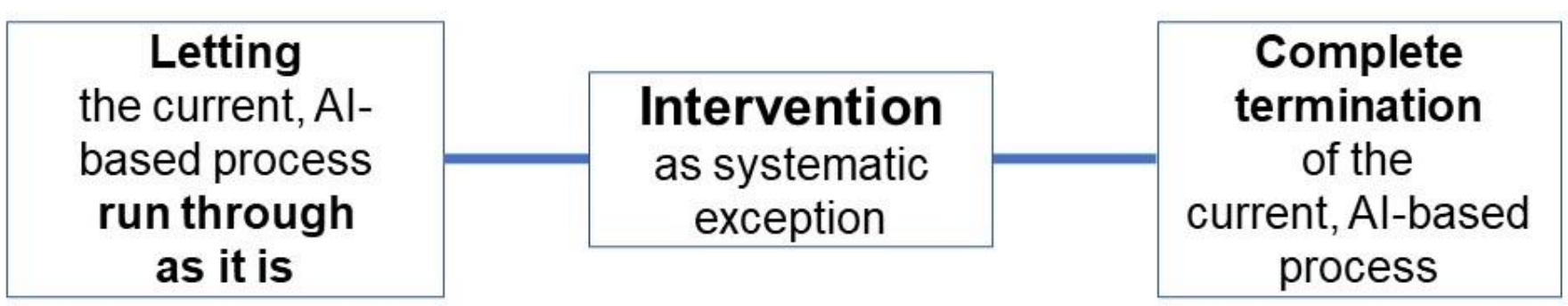

**Fig. 3.** Intervention – a third option between complete shutdown of a system and letting the current process run as it is

### 3.2 'Intervention into' vs. 'reconfiguration' of AI

Today's AI systems are rarely created in a single step and then used permanently in the same version. Instead, they are – for instance, in the case of machine learning – usually further trained on a regular basis to provide better predictions, decisions, and actions. In these cases, we speak of the AI system being reconfigured. Intervention, in contrast, is not a permanently effective reconfiguration of the technical system but an adjustment that is made exceptionally and temporarily. However, interventions can be an entry point for reconfiguration. Intervention should not be used to achieve repeatedly the same correction effect in similar situations. Intervention, then, as Figure 4 illustrates, is something third between two other poles, namely, on the one hand, allowing the process to run repeatedly always in the same form or, on the other hand, reconfiguring the technology-driven process with a permanent effect for the system itself. Intervention does not mean one or the other of these two poles. Instead, it opens a third option.

To understand this, it helps to use the simple, non-AI example of office coffee machines, which have a rotary knob that allows one to increase the amount of coffee powder or brewing water used beyond the preset level. That is an option for a single intervening activity that only applies to brewing one cup of coffee. Thus, an intervening activity is not an entirely new phenomenon. However, this rotary knob is not an example of a systematically designed intervention interface. By contrast, intervention interfaces must be implemented systematically, especially for AI applications.

According to the principle that intervention effects are limited to a certain time slot or a single case, if the knob on the coffee machine is turned, the quantities are only adjusted for a single brew; if this action results in a permanent adjustment of the quantities, it would be a reconfiguration. In the case of an intervention, the next coffee drinker can rely on the fact that the preconfigured quantity ratios are the same as always. Here, the socio-technical interplay of multiple users of a technical artifact highlights how critical it is that the effect of an intervention is limited. Whenever multiple people use AI applications, they must be confident that they will find the initial conditions as expected. This requirement is particularly important when it is too costly or undesirable to offer a personalized user interface to the participants. If one aimed at a systematic design for an intervention interface for the coffee machine, such a solution would allow users to decide whether their adjustment applies only for a single cup (intervention), for all further cups in the future (reconfiguration), or whether also previous adjustment quantities could be recalled. This simple example illustrates that an intervention interface opens the possibility for reconfiguration so that the design motivates users to fix the underlying reason for intervention instead of rejecting systems or repeating interventions again and again. If it becomes clear in the case of the coffee machine that almost all users increase the amount of brewing water, then this is a reason to think about permanent reconfiguration of the brewing water quantities.

A further example is the case of the brake assistant. If the intervention button is pressed every time a highway access is approached, this points towards the necessity of a reconfiguration. To this end, it must be possible to register and document interventions. AI could itself make suggestions as to when and how a reconfiguration

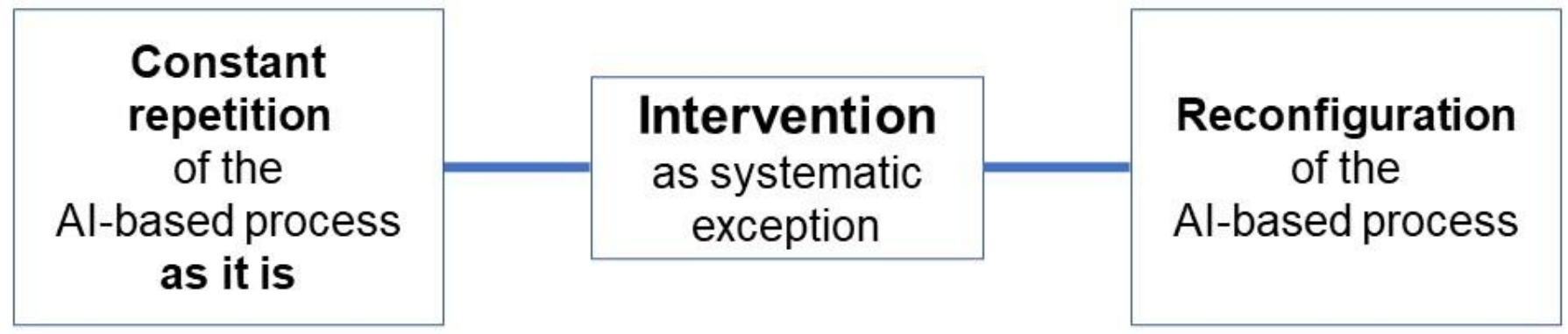

**Fig. 4.** Intervention – a third option between reconfiguration of a system and repeating an automated process as it is makes sense for the improvement of an AI system, so that, for example, by evaluating geoinformation at all highway accesses, the brake assistant is temporarily suppressed and at the same time a clearly perceptible warning signal is given in the vehicle cabin.

## 4 The integration of further development of AI and of human competencies

In summary, it can be said that a need to reconfigure the AI cannot necessarily be derived from rare interventions. So, if it only happens occasionally that a car parks near an unprotected tree area, then intervention is sufficient without the need to reconfigure anything. However, a permanent adjustment would make sense if the parking space were near one's home and used daily. Similarly, if almost all people with the same characteristic, such as dual citizenship, are systematically but inappropriately suspected of social fraud, then a reconfiguration is indicated. In the case of frequent, similar interventions, a permanent adjustment is appropriate. It then depends on the specific use case whether:

- the system can evaluate these interventions itself to suggest improvements or to give prompts that point towards possibilities of reconfiguration,
- the users make small adjustments by themselves or
- the users propose carrying out an adjustment and delegating this to technical experts or data analysts who are capable of reconfiguring the system and are responsible for it.

Any reconfiguration made must subsequently be transparent and explainable because it can also change the intervention effects and affect several people or organization units who use the system.

With respect to autonomy and oversight, it is the task of human actors to finally decide whether and how a reconfiguration takes place. Those who frequently intervene should help in this decision and can themselves make suggestions as to what kind of adjustment is appropriate. This requires that an intervention always be reflected upon, at least after it has taken place. To support this reflection, AI must be able to explain its reasoning mechanisms, as well as the data and influencing factors that are considered. Only when it is known where the deficits really lie – for example, at the level of sensor technology, the interpretation of data, automated decision-making, or in their implementation [35] – can meaningful adaptation measures be proposed. Interventions themselves and the effects that can be observed in the process help to better understand AI as well as the application domain in which AI is used. The weaknesses of the AI that are ironed out by an intervention help to think about the specifics of the workspace in which AI is used. Thus, interventions are a trigger for continuous learning on the job [36], [37]. With respect to fostering learning, even for a one-off intervention, it is advantageous to think about the facts the system should be aware of to avoid the necessity of the intervention. If users understand which underlying assumptions are implemented into the system and cause the need for an interven-tion, this helps them to better understand the characteristics of a domain and the shortcomings of models of this domain.

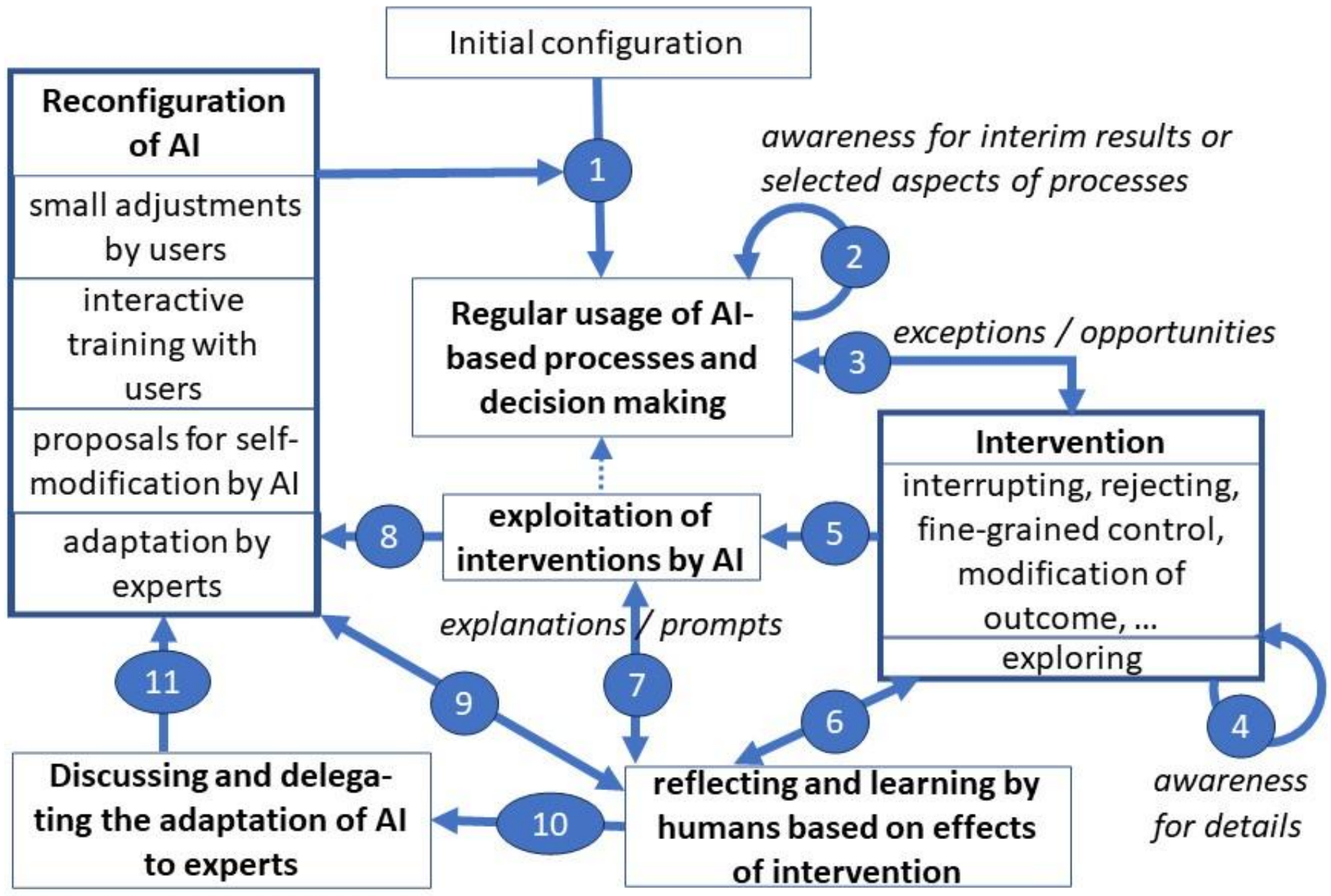

**Fig. 5.** Integrated further development of humans and AI

As can be seen from these considerations, intervention can result in learning progress for the users, which then also leads to further development of the AI in some cases. Figure 5 presents how the evolution of an AI system and learning on the users' side are generally intertwined by interventions. The relevant relations of Figure 5 can be described as follows:

1. Before regular use of AI-based processes and decision-making can take place, the system has to be appropriately configured. This can include phases of training with appropriate data, as is the case with machine learning, testing, selecting statistical methods to be applied, adding rules, and customizing the system with instructions that take the characteristics of users into account. After a period of usage, a reconfiguration might change the conditions of regular usage.
2. During regular usage, the users will not want or be able to be completely aware of all aspects of the AI-based process. This is particularly relevant if multiple processes run in parallel so that comprehensive surveillance would be mentally taxing and would overwhelm the users. Thus, users must decide on which AI outcome they focus on.

3. Based on their awareness of AI outcomes and processes, the users might detect that a situation exceptionally leads to inappropriate process control or decision-making by the AI system or that there is an opportunity to gain an advantage by adapting the behavior of AI. This is the starting point for an intervention in accordance with the options presented in the taxonomy (Figure 1). Usually, interventions are reasonable to alter AI outcomes. Additionally, or even alternatively, the users may want to intervene to check how a system reacts to modified data input, parameters, or rules. This is a kind of what-if exploration that aims to understand the relationships and key assumptions of the AI's model of a domain. The relation between intervention and regular usage is bi-directional since the users can at any time decide to return to regular use or even to revise the effects of an intervention.
4. As long as an intervention is active or exercised, the users must have a detailed, fine-grained awareness with respect to those aspects they want to influence by the intervention. Thus, if a driver decides to take control of the car by himself, he has to monitor the situation around the car. The more the intervention aims for exploration, the more effort is needed to observe the intervention's effects.
5. As illustrated in Figure 3, the AI system is not completely shut down, and an intervention does not entirely overthrow its results. By contrast, it remains active in the background and thus remains able to observe and record the intervention activities. This provides a basis on which the system can contribute to further improvement (see relation 9 and 7 in Figure 5). The configuration of the system might even specify that an intervention can be terminated by the system to return to regular use (see dotted line in Figure 5) if the intervention might lead into a dangerous situation.
6. The detailed awareness of what happens during an intervention triggers the users' reflection on the situation and on the outcome of their intervening activities. The more exploration is involved, the more learning can take place based on this reflection. The learning effects are not restricted to the features and characteristics of the AI system but also include a better understanding of the domain to which the model underlying the AI system refers. The reflection can also help to adjust the intervening activities.
7. AI might become active during the intervention to support human reflection during work [38] with the system. In a typical way of managing such a case, users ask for explanations to understand interim results, warnings, and decisions made by a navigation system. In contrast to this conventional way of seeking explanations, the AI system can take the initiative to explain actions or decisions against which a user intervenes. The AI system can also provide warnings, e.g., if certain assisting functions are interrupted. Moreover, the system could prompt users on how to act during an intervention or on how to proceed after the intervention. Further development of AI should allow providing prompts that describe how the system could be adapted so that it reacts adequately by itself in situations that have led to interventions.
8. Instead of prompting the users to carry out an adaptation, the system can outline proposals on how it could be adequately modified and document them within the system itself. These proposals could be reviewed by technical experts (see relation 11, Figure 5) who can instantiate the modification or adapt it. Furthermore, the data that is recorded during interventions can be used as examples with which a machine learning system can retrain its model. Thus, " … constant performance improvement is attainable by continuous training by users by providing the system with feedback that indicates misclassifications …"[39, p. 440]. This type of evolving is considered interactive machine learning [40]. It might be specified that changes in the AI's domain model, which are based on retraining, must be agreed upon by a human expert before they are implemented (see relation 11, Figure 5).

9. Besides self-adaptation by the AI system, users can also contribute modifications based on their reflections. These might be reasonable for small adjustments by specifying, for example, that the modification of thresholds or rules should be made permanent. This relation is bidirectional since users should be permanently informed if the system is reconfigured.
10. Reflecting on the effects of intervention can lead to discussing issues about the AI system with other users, user communities, or experts who are familiar with the AI system. These discussions can initiate further learning about the system and the domain it covers. In particular, the discussion can complement AI-based explanations with human-based explanations. This kind of hybrid XAI provides a broader picture of the capabilities and limits of a system. Eventually, experts might be able to propose more sophisticated adaptations that contribute to a reconfiguration of the system.
11. The discussions might eventually lead to the implementation of adaptation proposals that are either provided by the AI system itself or by experts. Overall, it is reasonable to include human expertise before a reconfiguration is completed. It can be considered as an outcome of managerial decisions and organizational practices on how the reconfiguration is coordinated and approved. Such coordination (see Figure 6) is highly relevant since AI usage is part of a socio-technical system where several stakeholders are affected.

In conclusion, the role of intervenability and intervention interfaces, as presented in Figure 5, helps to meet the requirements that AI is "…designed and trained to augment individual workers, and augmentation involves a mutual learning process through which both humans and AI learn from each other and hence coevolve" [41, p. 91].

## 5 Basic requirements for designing the intervention interface

Intervenability is a design criterion that relates to the interaction between humans and AI when processing tasks. For intervenability to be successful, the interaction interface to the AI must meet several requirements. Up to now, no empirical studies have derived such requirements. In their paper on intervention interfaces, Schmidt and Herrmann [1] refer to Shneiderman's [42] golden rules to derive requirements for intervenability, including six design principles. In what follows, we point out some crucial requirements that are important for designing intervention interfaces.

- Information must be offered or easily retrievable that can be used to decide whether and how to intervene and realize the intervention's possible effects. Accordingly, all AI usage should be accompanied by informative feedback [42]. Furthermore, users should not be required to keep anything about the system's previous status in mind to handle and terminate intervention successfully. This corresponds to "reduce the short-term and working memory load" [43, p. 5]. A well-developed AI system could even suggest possible steps for intervention or for improving the system itself (see Figure 5).
- It must be comprehensible when the intervention begins and ends. In every situation –before, during, or after intervention – the intervention interface must clearly communicate how responsibilities and control are shared between the human and the system [1, p. 45]. This kind of comprehensibility matters not only to those who start the intervention but to everyone who is affected by the change during AI usage. This requirement can be derived from a socio-technical view and is particularly relevant if several processes and roles [44] are involved, as can be seen in the context of the smart home [45].
- The operating elements with which one controls intervention must be easily and quickly accessible and easy to use – just like the emergency stop button [46]. Similarly, intervention might be started by a physical button. As we know from the equipment of smart homes, various control elements such as smartphones, tablets, or computers [22] can be used to trigger interventions. This kind of solution for intervention interfaces provides more flexibility so that several options can be offered to intervene, as presented in the taxonomy (Figure 1).
- Intervention must be quickly terminable, and it must be possible to undo the effects of intervention if necessary. This is an important prerequisite if interventions are employed to explore what-if questions by altering the parameters of the AI system. The intervention interface should offer a simple means to reverse the impact of the system's automated behavior, e.g., to cancel an alarm, or undo the results of interventions themselves [1, p. 45].
- Interventions and their effects must be documented so that they can be reflected afterward with respect to their appropriateness and can be used as a basis for further adjustments to the AI system. In working practice, it will be appropriate that the employees leave comments in the system about the reasons for which one intervenes. This requirement can be derived from the cyclic integration of intervention and AI reconfiguration, as presented in Figure 5.

- Intervention interfaces must allow the user to choose among the various options listed in the intervention taxonomy (see Figure 1) as far as they are appropriate for the application domain. For example, an intervention must contribute to achieving certain effects immediately, as well as effects that lie in the future.
- The performance of AI, as well as its shortcomings, can take place on different levels, such as sensing, deciding, and acting [35], [47]. Sensing includes sensor-based data acquisition, deciding also covers the interpretation and planning based on the data, and acting refers to all kinds of AI behavior that renders effects in its environment. The difference between these levels has been illustrated above by the example of assisted parking. Intervention interfaces should help the users to selectively address these levels appropriately to minimize their effort. This requirement can be derived from the socio-technical principle of "balance of effort and experienced benefit …" [28, p. 3101].

Fulfilling these requirements helps to keep people involved in the use of AI, which corresponds to the principle of oversight, autonomy and keeping humans in the loop.

## 6 Organizational practices to support intervenability

Keeping humans in the loop can be sustained only if organizational practices permit and promote its implementation. Herrmann and Pfeiffer [48] use the example of predictive maintenance to argue that the human-in-the-loop principle must be supplemented by the principle of 'keeping the organization in the loop'. It is obvious that many employees will only dare to reject an AI result – for example, not to follow a maintenance request – if their organizational practices safeguard such a decision or even promote it. We conceive organizational practices as interactions and negotiations within an organization that are based on structures and processes that are subject to their own logic [49], [50].

Following a diagram of the organizational embeddedness of AI, where Herrmann and Pfeiffer [48, p. 1537] present several interacting organizational practices, Figure 6 depicts eight aspects of reciprocal influence that are relevant to bringing intervention into reality:

1. In the center of Figure 6 is the managerial task of offering and coordinating the possibilities for intervention into AI based processes, and for the evolution of AI. This coordination is important in organizational units since AI is part of a socio-technical system where changing a regular process or decision-making can influence others. This may include risks if others do not understand whether an intervention is still active or is already terminated. Thus, managerial coordination provides rules that guide the operational forces by clarifying what kind of intervention can be initiated by whom under which conditions. The handling of the original tasks that are supported by employing AI must be organized in a way that promotes intervention. The rules for coordinating the handling of AI might be negotiated, adapted, and eventually adopted through organizational practices. Furthermore, the coordination must specify how the interplay between intervention into and evolution of AI (reciprocal influence 7 in Figure 6), as presented in Figure 5, is implemented and which role the technical team has to take over for reconfiguring AI.

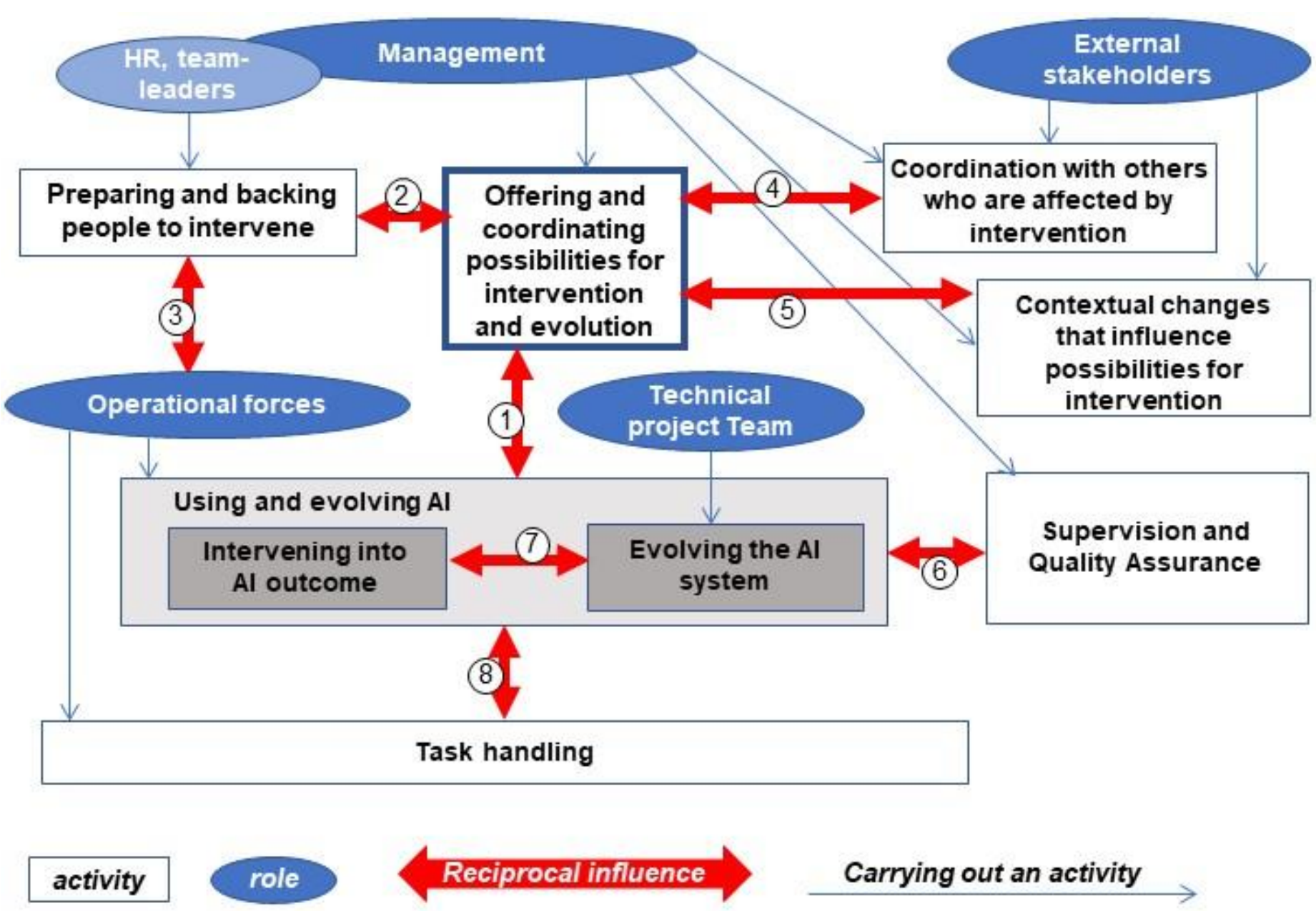

**Fig. 6.** Managerial tasks and organizational practices in the context of intervenability

2. Making intervention happen requires not only that people be allowed to intervene. They also need to be prepared and encouraged to do so. Thus, the coordinative measures must be coupled with HR development and supported by team leaders so that flexible handling of AI results takes place, such as rejecting or adjusting them by intervention. It must also be part of organizational practice that both AI's and employees' capabilities are continuously developed and that everyone is encouraged to contribute to this learning process.
3. Preparing, encouraging, and backing the operational forces to conduct interventions requires an intensive exchange with them. Interventions, as well as shortcomings of AI, can be documented to provide a basis for reflection that should not only take place individually but also collaboratively within teams and with supervisors. Employees need to know that their peers and supervisors will recognize it when they intervene to counteract or test the performance of the AI.
4. Since organizational units interact with other stakeholders, they also must be informed about the possibility of intervention that might take place and affect them. It must be specified who in the environment will be aware of interventions that took place, whose interests might be affected, and how the reaction of others has to be taken into account when an intervention is instantiated. For example, if an AI-based smart home triggers an alarm and the owner overrides this alarm, then security services must be able to verify that this was the result of an intervention by the inhabitants of the smart home.

5. The environment of an AI application underlying continuous changes must be considered by management when coordinating the possibilities for intervention. Changes may be triggered by new technologies or by the market. Changes in contracts or laws might require standard procedures that prevent certain types of interventions that violate the regulations. However, interventions might even be necessary to allow for a quick reaction to meet the requirements of new regulations or ethical discourses about AI. Besides shortcomings of a certain AI application, the continuous change of the context of AI usage is a further trigger that makes interventions and continuous evolution of AI necessary.
6. From an organizational point of view, interventions into AI, its evolution, and the interplay between both need close supervision. This includes testing AI and its reconfiguration, evaluating whether interventions were reasonable or not, and assessing the configuration of the intervention possibilities itself, for example, whether the time that elapses before the system returns to the regular procedures is too long or too short. In particular, it is a management task to consider whether the same kind of intervention happens too often and should be avoided by reconfiguration.
7. As described in Figure 5, the intervention in and evolution of AI are intertwined. Interventions influence how an AI system will evolve or be reconfigured, and the reconfiguration itself changes the possibilities and conditions of intervening. These changes must be regarded by the organizational practices that surround interventions.
8. Using and influencing AI must be closely related to the task handling that AI should support. The task-handling procedures might have to be adapted and adjusted to AI usage. For example, employees can be encouraged to first consider for themselves what decision they would make before looking at the AI's proposed decision [51]. Such a measure makes reflection on AI results and their possible rejection more likely and contributes to the calibration of trust in AI.

The reciprocal influences of Figure 6 illustrate that intervenability is a criterion that must be implemented within a holistic, socio-technical approach that integrates organizational practices, individual activities and capabilities, technical artifacts and their evolution, as well as the development of human competencies [52].

## 7 Conclusion

Based on the literature and several examples of possible AI applications, we outlined the concept of intervenability. It is a new phenomenon that is not covered by emergency shutdown, workarounds, or reconfiguration of automated systems. Intervenability instantiates the principles of controllability, autonomy, oversight, and keeping humans in the loop in the context of AI. We provide a taxonomy that includes several categories of intervening activities and differentiates them with respect to the user's mental effort. This taxonomy goes beyond the initially proposed concept for the intervention of Schmidt and Herrmann [1] by extending its scope of applicability from real-time control of automated processes to AI-based discrete case-related decision-making. This is in accordance with human-centered AI, which seeks to employ human strengths combined with the usage of AI. We demonstrate how intervenability can potentially contribute to the ongoing development of human capabilities on the one hand and to further improvement by reconfiguring AI on the other. Exploring and collaboratively reflecting on the effects of interventions is key to enabling this ongoing improvement on both sides. Intervenability also provides further momentum for the design of an AI that can, by itself, help realize interventions and advance a smooth transition from intervention to reconfiguration of the AI.

The presented concept of intervenability and intervention interfaces faces some limitations. So far, we can find single instances of intervention possibilities, such as the party button for heating, but no systematically designed complete intervention interfaces. Thus, it is difficult to empirically investigate the appropriateness of intervenability. Furthermore, conducting experimental lab studies is difficult since intervenability is a solution that reacts to problems that occur only exceptionally and cannot be anticipated or systematically reproduced. Furthermore, the limits and benefits of possibilities for intervention can only be studied within a socio-technical context where technology, human competencies, and organizational practices need to be appropriately developed and closely intertwined. We have demonstrated (Figure 6) how managerial measures of coordination and the adoption by organizational practices must support interventions. It is not sufficient to allow workers to intervene in AI outcomes or processes, but they also must be made competent and encouraged to do so. Intervention must not be a concern only of individuals but of teams. Fundamentally, intervenability must be implemented in such a way that the benefits gained from an intervention outweigh the effort involved.

These limitations point to future challenges. From an empirical point of view, examples of the relevance of intervenability need to be identified and contrasted with the concept presented above. We will have to explore how intervenability can serve as an umbrella for various ways of enabling human autonomy, oversight, and controllability when using AI. Empirical cases will be needed to show how the use of systematically designed intervention interfaces can be beneficial. A systematic, socio-technically oriented approach could stimulate further research into how interfaces can be designed for the use of AI – especially if several autonomous processes run in parallel – and how AI can be implemented organizationally.

**Funding and Acknowledgements:** This work was supported by the project Humaine (Human-centered AI Network), which is funded by the Federal Ministry of Education and Research (BMBF), Germany, within the '*Zukunft der Wertschöpfung – Forschung zu Produktion, Dienstleistung und Arbeit*' Program (funding-number: 02L19C200). We thank Christopher Lentzsch, with whom we have intensively discussed the socio-technical enlargement of intervention concept and who contributed many suggestions to the intervention taxonomy and the design of the intervention button (see Figure 2).